# Real-time observation of non-equilibrium phonon-electron energy and angular momentum flow in laser-heated nickel

Vishal Shokeen[1], Michael Heber[2], Dmytro Kutnyakhov[2], Xiaocui Wang[1], Alexander Yaroslavtsev[1], Pablo Maldonado[1], Marco Berritta[1], Nils Wind[3,4], Lukas Wenthaus[2], Federico Pressacco[2], Chul-Hee Min[4,5], Matz Nissen[3], Sanjoy K. Mahatha[4], Siarhei Dziarzhytski[2], Peter M. Oppeneer[1], Kai Rossnagel[4,5], Hans-Joachim Elmers[6], Gerd Schönhense[6], Hermann A. Dürr[1*]

[1] Department of Physics and Astronomy, Uppsala University, 751 20 Uppsala, Sweden
[2] Deutsches Elektronen-Synchrotron DESY, 22607 Hamburg, Germany
[3] Institut für Experimentalphysik, Universität Hamburg, 22761 Hamburg, Germany
[4] Ruprecht Haensel Laboratory, Deutsches Elektronen-Synchrotron DESY, 22607 Hamburg, Germany
[5] Institut für Experimentelle und Angewandte Physik, Christian-Albrechts-Universität zu Kiel, 24098 Kiel, Germany
[6] Institut für Physik, Johannes Gutenberg-Universität Mainz, 55128 Mainz, Germany

*Corresponding author. E-mail: hermann.durr@physics.uu.se

**Abstract**

Identifying the microscopic nature of non-equilibrium energy transfer mechanisms among electronic, spin and lattice degrees of freedom is central for understanding ultrafast phenomena such as manipulating magnetism on the femtosecond timescale. Here we use time and angle-resolved photoemission spectroscopy to go beyond the often-employed ensemble-averaged view of non-equilibrium dynamics in terms of quasiparticle temperature evolutions. We show for ferromagnetic Ni that the non-equilibrium electron and spin dynamics display pronounced variations with electron momentum whereas the magnetic exchange interaction remains isotropic. This highlights the influence of lattice-mediated scattering processes and opens a pathway towards unraveling the still elusive microscopic mechanism of spin-lattice angular momentum transfer.

**Introduction**

The ability to drive a system far out of equilibrium by absorbing a femtosecond (fs) laser pulse provides access to novel dynamical phenomena *(1-6)*. Such far-from-equilibrium dynamics offers non-thermodynamic pathways for the ultrafast control of materials' properties before thermalization on longer timescales is reached *(7, 8)*. While the initial ultrafast laser heating of the electronic system and the subsequent energy transfer to other degrees of freedom (spin, lattice) can be studied in so-called pump-probe measurements *(9-14)*, the understanding of such strongly non-equilibrium dynamics in solids is still very limited. Relying on the quasiparticle description of electronic, spin and lattice excitations, temperatures are often assigned to the individual quasiparticle reservoirs *(15)*. It is assumed that the quasiparticle subsystems are each in separate equilibrium at all times and reach global equilibrium by exchanging heat *(15, 16)*.

The ensemble-averaged non-equilibrium heat exchange between electron, spin and lattice quasiparticles can be described by a rate equation treatment within the 3-temperature model which leads to different demagnetization timescales depending on pump fluence *(10)*. However, it has so far been difficult to identify the quasiparticle states that are directly involved in energy, momentum and, for magnetic systems, angular momentum transfer processes. Several attempts have been made to identify points in energy and momentum space that facilitate electronic spin-flip processes deemed necessary for ultrafast demagnetization of ferromagnets *(17, 18)*. It remains ambiguous which phonons take up spin angular momentum in a non-equilibrium demagnetization process *(19, 20)*. While the excited electron system reaches rapidly a uniform electron temperature through fast electron-electron scattering within a few hundred femtoseconds *(11, 21)*, phonon equilibration can take tens of picoseconds *(22-25)*. In addition, energy transferred from electrons to certain phonon modes can be donated back to the electrons much faster *(24, 25)* thus increasing the potential complexity of non-equilibrium pathways towards final equilibrium.

Here we aim to identify the quasiparticle states responsible for energy and angular momentum redistribution between the electronic system and the crystal lattice. We expect that this provides the experimental benchmarks for modeling and ultimately understanding the enigmatic ultrafast spin-lattice angular momentum transfer process. Traditionally this has been attempted by following the energy transfer from the electronic system to the lattice. However, in this case, many different scattering processes between electrons, spins and lattice excitations prevent the unambiguous identification so far and mainly ensemble-averaged information via the 3-temperature model is available. In the present paper we approach this topic from a different angle. Based on our previous observation that non-equilibrium phonon modes are populated first, we observe in real time how these phonon modes scatter with valence electrons and affect their magnetic moments. We show that these non-equilibrium scattering processes can be unambiguously separated for the ensemble-averaged dynamics.

Our approach is depicted in Fig. 1. During the first picosecond after fs laser heating phonon modes at the Brillouin-zone (BZ) boundary become populated first while the lower-energy BZ center phonon modes become populated only on much later times *(24)*. When the energy stored in the non-equilibrium population of BZ-boundary phonon modes exceeds that of the electronic system a back-transfer becomes energetically favorable as displayed in Fig. 1A *(24, 25)*. We use time- and angle-resolved photoemission spectroscopy (tr-ARPES) to directly identify the underlying energy back-transfer processes. These are based on absorbing a non-equilibrium zone-boundary phonon (Fig. 1B) with momentum, $\mathbf{q}_p$, and energy, $E_p$, (see methods) while scattering electrons from occupied (momentum $\mathbf{k}_e$ and energy $E_e$) to unoccupied states ($\mathbf{k}_e$' and $E_e$') by conserving momentum ($\mathbf{k}_e + \mathbf{q}_p, = \mathbf{k}_e$') and energy ($E_e + E_p = E_e$') as shown in Figs. 1C, D.

Relevant BZ-boundary phonons in Ni have energies of typically 25-30 meV *(24)* and, therefore, we expect to see differences in band-occupations on such energy scales and for electronic momentum values that are separated by a BZ-boundary phonon momentum. We report tr-ARPES measurements that detect such changes in the band-filling of laser-heated Ni using radiation from the FLASH free electron laser to cover the complete Ni Brillouin zone. Different electron band occupations are quantified by a transient, electronic state-dependent chemical potential. We identify several pairs of Bloch states, separated by the momentum of a zone-boundary phonon, whose transient chemical potentials differ approximately by the phonon energy. This provides the direct visualization of p-e scattering processes in real time. In addition, we find a surprising fluence dependence of this process that points to a competition between electron-phonon energy transfer. We show that this has direct consequences for the state-resolved magnetic moment dynamics deviating from the generally assumed global dynamics averaged over the whole ensemble.

**Results**

We measured tr-ARPES from 2 and 9 nm thick Ni(111) films grown on W(110) at the FLASH free electron laser in Hamburg (see methods). Our main goal is to identify the p-e scattering processes leading to energy and angular momentum redistribution between the phonon and the electronic system. To achieve this, we need to identify $k$-points in the Ni BZ that satisfy the characteristics described in Fig. 1 for non-equilibrium phonon-electron energy transfer processes. The Ni electronic structure is characterized by itinerant 4$sp$-states whose energy disperses strongly with electron momentum and 3$d$-bands with flat energy dispersions. The latter are split by the exchange interaction into fully occupied Bloch-states of majority spin character and partially occupied minority spin-states. The difference in occupation numbers represents the size of the Ni magnetic moment. Ni 4$sp$-bands display a spin-splitting only via hybridization with the 3$d$-states. Ni is thought to demagnetize both in equilibrium and following ultrafast laser excitation by a reduction of the 3$d$ exchange splitting and a corresponding redistribution of the electron occupation between majority and minority states *(11, 21, 26-29)*.

We study electronic bands in the L-W-U and Γ-K planes shown in Fig. 1 that are separated by zone-boundary phonon wavevectors. We include representative ARPES datasets and band-structure calculations in the supplementary Fig. S1. The Ni electronic structure of the L-W-U plane is largely characterized by exchange-split 3$d$-bands straddling the Fermi level, $E_F$. Only near the L-point are itinerant $sp$-bands hybridizing with the magnetic 3$d$-states. The situation is different in the Γ-K plane where a large proportion of the bandstructure is characterized by $sp$-band crossing the Fermi level. Only along the $\Gamma - d_{11\bar{2}}$ direction we find strongly dispersing 3$d$-bands and the minority-spin 3$d$-band crossing the Fermi level in agreement with previous studies *(11, 21, 26-29)*. We note that $d_{11\bar{2}}$ is not a high-symmetry point. Instead, it describes a minority-spin hole-pocket located approximately halfway between X and L points *(29)*. It is worth noting that the crossing points between $sp$ and 3$d$-states can contain so-called spin-orbit hotspots, i.e. points where spin-orbit coupling mixes majority and minority spin character. It is thought that these hotspots play a crucial role in non-equilibrium angular momentum exchange between the electronic system and the lattice *(17, 18)*.

Figure 2 (extended data sets are shown in Figs. S2-4) shows the case of the $sp$-band Fermi-crossing along the Γ-K direction. The tr-ARPES data were evaluated for the ($k_x$, $k_y$) regions shown in Fig. 2C. Since the $sp$-bands have a steep dispersion, we shifted the $k$-integration regions for different $E$-$E_F$ values according to the observed dispersion of

$\Delta E/\Delta k = 1.65$ eVÅ (see Fig. 2D). We note that the *k*-integration region has been chosen wide enough to average over the exchange-split majority and minority spin *sp*-bands. The data in Fig. 2 can then be fitted by a Fermi-Dirac distribution function with the electronic temperature, $T_e$, and the chemical potential, $\mu$, as parameters (see supplementary information). We assume a constant *sp*-band density-of-states for the $E-E_F$ values in Fig. 2. The resulting delay-time dependencies of the fit parameters will be discussed below. Here it is important to highlight the clearly visible chemical potential shift. This effect can be unambiguously separated from the observed broadening due to increased electronic temperatures, $T_e$, after optical excitation *(11, 21, 28)*. We also point out that the chemical potential shift displays an interesting fluence dependence that is contrary to expectations, i.e. the chemical potential shift is larger at 2.2 mJ/cm² (Fig. 2A) than at 3.7 mJ/cm² (Fig. 2B). We will discuss this behavior below as a competition between p-e energy transfer and electron-electron scattering.

The situation is more complex for the 3*d*-bands shown in Fig. 3, where also the change in exchange-splitting and spin-dependent band-occupation needs to be taken into account. Figures 3 A-D show the dynamics at the W-point (see Fig. 1B and inset of Fig. 3C for the position in momentum space). We model the exchange-split 3*d*-bands as illustrated in Fig. 3A before and in Fig. 3B 1.8 ps after a 3.7 mJ/cm² pump pulse. Ultrafast demagnetization *(9)* is approximated by asymmetrical energy shifts where majority spin states below $E_F$ move twice as fast towards $E_F$ than minority spin states located above $E_F$ *(27)*. This leads to an excellent match (dashed lines) with the measured tr-ARPES data (symbols) (Fig. 3C) considering a time-independent background *(21, 30)* (dotted line). Figure 3D illustrates the reason for the observed spectral changes. The intensity reduction observed near $E-E_F = -0.3$ eV reflects the shift of the occupied majority spin state towards $E_F$. Consequently, it becomes partially depopulated as shown by the (green) shaded areas beneath the majority peak (solid green line) in Figs. 3A, B. The increase of intensity near $E_F$ is mainly due to the population transfer (yellow shaded area) from the majority to the minority-spin states shifted to lower energy (yellow solid line). At the W-point shown in Figs. 3A-D these processes are largely unaffected by a change in chemical potential as indicated by the gray dash-dotted line ($\Delta\mu = 0$ meV) being nearly identical to the green dashed line ($\Delta\mu = 20$ meV) in Fig. 3D. We used electronic temperatures that were obtained from data shown in Fig. 2 measured at identical pump fluences.

In contrast, the tr-ARPES results at the $d_{11\bar{2}}$-point (see Fig. 1B and inset of Fig. 3G for the position in momentum space) shown in Figs. 3E-H display characteristic differences. While the changes in exchange splitting are identical to those at other *k*-points (see Fig. S6), the chemical potential shifts by up to $\Delta\mu = 60$ meV as illustrated by the Fermi-Dirac distribution curves (purple dashed lines) in Figs. 3E, F. This results in a negligible intensity reduction near $E-E_F = -0.3$ eV and to a stronger intensity increase at $E_F$ in Fig. 3H. The latter is strongly affected by the energy position of the chemical potential (see green dashed and gray dash-dotted lines in Fig. 3H). We find similar characteristics as for the W and $d_{11\bar{2}}$-points in Fig. 3 for all analyzed *k*-points (see Figs. S2 and S3 for 3.7 and 2.2 mJ/cm² pump fluences, respectively). The resulting *k*-dependent values for the delay time dependence of the local chemical potential, electronic temperature, exchange splitting and spin-dependent band occupation are summarized in Fig. 4 and represent the central result of this paper.

**Discussion**

The results shown in Figs. 4A, E reflect the commonly accepted ultrafast dynamics that occur in ferromagnetic Ni *(9, 11, 21, 28)*. Laser excitation deposits energy into the electronic system, which upon thermalization leads to an increase of the electronic temperature, $T_e$. We find maximum $T_e$ values of 840±20 K and 700±30 K for pump fluences of 3.7 mJ/cm$^2$ (Fig. 4A) and 2.2 mJ/cm$^2$ (Fig. 4E), respectively. This leads to a demagnetization via transfer of energy and angular momentum to the lattice *(19, 20)* and excitation of magnons *(21, 31)*. The observed changes in Fig. 3 (see also Fig. S6) of the majority-spin-band binding-energy reflects the ultrafast collapse of the magnetic exchange splitting *(11, 21, 28, 32)* and an associated repopulation of majority/minority spin bands (see Fig. 3) that is believed to be responsible for ultrafast demagnetization in Ni *(10)*.

Here we report the observation of a momentum-dependent change in band-filling that we describe as the build-up of quasi chemical potentials. Spin-integrated results are summarized in Figs. 4B, C and F, G for pump fluences of 3.7 mJ/cm$^2$ and 2.2 mJ/cm$^2$, respectively. The build-up of such chemical potential changes is slightly delayed compared to the increase in electronic temperature and to the energy shift of the majority-spin 3$d$-bands. The changes in the momentum-dependent chemical potential also persist up to the longest time-delays measured, outlasting in particular the picosecond timescale remagnetization and electronic temperature decay, thus, indicating a different physical origin.

Figures 4D, H display the dynamics of momentum-dependent magnetic moments, $\Delta M_k$, at the respective $k$-points. Extended datasets are shown in Fig. S14. $\Delta M_k$ was evaluated from the band occupations $N_{\text{maj, min}}$ of majority and minority spin bands in Fig. 3 as $\frac{\Delta M_k}{M_{0,k}} = \frac{N_{maj} - N_{min}}{N_{0,maj} - N_{0,min}}$, where $M_0$ and $N_0$ represent the magnetic moments and occupation numbers, respectively, before laser excitation. The results displayed in Figs. 4D, H and S14 clearly show that the magnetization dynamics at the investigated $k$-points is non-uniform, i.e. different $k$-points contribute differently to the total magnetic moment changes. This is in contrast with the observed changes in magnetic exchange splitting (Fig. S6) that is the same for all measured $k$-points. We will discuss below that the observed changes are compatible with a spin-dependent energy transfer between lattice and electronic system.

It has been argued that ultrafast demagnetization of Ni is accompanied by a valence-band narrowing observed in transient 2$p$-3$d$ x-ray absorption *(33, 34)* and ascribed to pump-induced non-equilibrium electron populations (35*)*. This has more recently been confirmed experimentally and theoretically described as ultrafast changes of electronic correlations *(36)*. Such changes in the Ni valence bands could possibly influence the observed chemical potential shifts. We have, therefore, probed the influence of both effects on Ni 3$p$ core levels for 3.7 mJ/cm$^2$ (see Fig. S7). We observed a shift of the 3$p$ core-levels toward the Fermi level by 17±3 meV, which is very close to the chemical potential shift of $d$-bands at the U and W points. However, the observed core-level shift is smaller than the chemical potential changes observed for *sp*-bands and at the $d_{11\bar{2}}$-point. Since core-levels probe changes of the electronic system averaged over of the whole valence-band *(37)*, valence-band narrowing related to the collapsed 3*d* exchange splitting and electronic correlations cannot explain the observed chemical potential changes in Fig. 4.

The enhanced chemical potentials of the *sp*-bands and at the $d_{11\bar{2}}$-point indicate that more electrons reside in these states for delay times longer than ~0.5 ps (see Fig. 4). It is a possible option that these extra electrons are photoexcited in the W(110) substrate and reach the surface of our 2 nm thick Ni(111) films via superdiffusive electronic currents *(38)* since also part of the W substrate is heated by the pump laser pulses. We performed additional experiments for 9 nm thick films (see Fig. S8) and found nearly identical chemical potential changes compared to the 2 nm films for the same absorbed pump fluences, i.e., for conditions leading to similar electronic temperature rises. Since the grazing incidence pump laser penetration depth is around 13 nm *(21)*, we expect less heating of the W substrate for the thicker 9 nm Ni film. Consequently, less electrons would accumulate and be visible in surface sensitive tr-ARPES than for the thinner 2 nm film under otherwise identical conditions. This observation clearly rules out the effect of superdiffusion for any surface accumulation of electrons possibly resulting in an electronic state-dependent band filling. In addition, we would expect purely electronic transport phenomena to evolve in a similar way to the electronic temperature, which is clearly not the case for the chemical potential evolution shown in Fig. 4.

The observed step-like chemical potential changes, *i.e.*, a sudden rise up to ~0.5 ps and a nearly constant value thereafter, require a non-electronic driving force to maintain the observed population imbalance within the Ni Brillouin zone. Hot non-equilibrium phonons could be such a force counteracting the equilibrating effect of an elevated electronic temperature, i.e. electron-electron scattering processes. Studies of non-equilibrium phonon dynamics show that Brillouin-zone boundary phonon modes are preferentially populated and essentially retain their energy for all times considered in the present paper *(24)*. This is the scenario depicted in Fig. 1. In the following, we argue that we have identified two scattering channels through which energy can be transferred back to the electronic system.

Electronic states at the W and $d_{11\bar{2}}$-points as well as at the U-point and for *sp*-bands are separated in momentum space by zone-boundary phonon momenta, $\mathbf{q}_p$ (purple arrow in Fig. 1B). Absorption of a phonon with momentum $\mathbf{q}_p$ can scatter electrons from states at W to $d_{11\bar{2}}$-points and from U to *sp*-bands. Since our system is far from equilibrium after the pump pulse, a significant number of such scattering processes can take place and consequently electronic states at $d_{11\bar{2}}$ and *sp*-bands are populated up to an energy of the phonon energy $E_p$ of about 25-30 meV *(24)* averaged over the ensemble of scattered electrons. This over-population agrees remarkably well with the observed chemical potential differences shown in Fig. 4. We note that the changes in band-filling are smaller in Ni than in other materials *(39, 40)* due to the smaller phonon energies.

Interestingly, the observed chemical potential changes shown in Fig. 4 scale inversely with the pump fluences, *i.e.,* they are larger at 2.2 mJ/cm² as compared to 3.7 mJ/cm². This points to a competing mechanism that counteracts the chemical potential build-up via phonon scattering. This may be due to the higher electronic temperatures at larger pump fluences, leading to more electronic scattering processes that can equilibrate population differences throughout the Brillouin zone. We can provide additional evidence for such a competition by investigating the involved timescales. While we do not obtain phonon-electron scattering times, $\tau_{\text{e-p}}$, directly in our experiments, it is possible to estimate them from high-resolution ARPES experiments *(29)*. In Ni so-called kinks due to electron-phonon interactions have been observed near the Fermi level for minority-spin 3*d*-bands and from the measured lifetime broadening *(29)* we can estimate $\tau_{\text{e-p}} \sim 100$ fs. The competing process would be electron-electron scattering that leads to an increased lifetime broadening as observed in our tr-ARPES measurements. From the fitted 3*d*-bandwidths at various *k*-points (see Fig. S11) we obtain lifetime broadening increases of 20-40 meV and 40-50 meV for fluences of 2.2 and 3.7 mJ/cm², respectively. These values correspond to electronic scattering times, $\tau_{\text{e-e}}$, of approximately 15 fs, i.e. they are much faster than the typical phonon-electron scattering times. It is, therefore, possible that phonon-electron energy transfer processes can be interrupted by the faster electronic scattering events, especially at higher fluences when more electronic scattering takes place. Furthermore, the BZ-boundary modes are the first phonons that become populated *(24)* and are very likely the reservoirs of angular momentum transferred from the spin system during ultrafast demagnetization *(20)*.

The changes in momentum-dependent magnetic moments observed in Fig. 4D, H demonstrate that the p-e energy transfer is also spin dependent. Demagnetization signified by a reduction of $\Delta M_k$ corresponds to electronic spin-flips that increase the minority band occupation at the cost of that of majority spin-states. For a pump fluence of 3.7 mJ/cm² $\Delta M_k$ at the $d_{11\bar{2}}$-point largely follows those at W and U. However, the situation is very different at a lower fluence of 2.2 mJ/cm² where the $d_{11\bar{2}}$-point dynamics remains mainly unaltered compared to the higher fluence but the dynamics at W and U are considerably smaller. The strong demagnetization at $d_{11\bar{2}}$ requires a transfer of minority spins from other *k*-points such as W and U. This is only possible via the uninterrupted p-e scattering described above.

Our results show the transient buildup of electronic populations in bands at different points in the Brillouin zone following ultrafast demagnetization of ferromagnetic Ni. These effects have been overlooked in previous tr-ARPES studies *(11, 21, 28)*, but are unambiguously separated from possible pump-induced space charge effects by probing many different momentum points simultaneously using momentum microscopy at suitably high photon energy that allows covering a substantial part of the Ni Brillouin zone. Our results establish the previously proposed *(24)* energy back-transfer from transiently excited zone-boundary phonons to the electron system to be spin dependent. Pump fluence-dependent measurements indicate that this energy back-transfer competes with another channel, possibly caused by quasiparticle scattering within the electronic system. The latter is ultimately responsible for achieving thermal equilibrium after transient phonon populations have decayed. We expect these energy relaxation channels to be operative in most materials under non-equilibrium conditions.

## Materials and Methods

### Ni sample growth and characterization.

Ni(111) films of 2 and 9 nm thickness were grown on a W(110) single crystal by molecular beam epitaxy using e-beam evaporators at a base pressure of $5 \cdot 10^{-10}$ mbar. The W(110) substrate was cleaned via several cycles of annealing (1200°C) in the presence of oxygen ($1 \cdot 10^{-7}$ mbar) followed by flashing to 2200°C. The deposition of nickel was initiated right after the high-temperature flash, so the substrate was cooling down during the growth of first Ni layers. The deposited Ni(111)/W(110) films were post-annealed at 400°C for a duration of ~5 min and quality of the deposited films was checked with LEED. New films were deposited every ~24 hours or earlier in case surface contamination could be detected. All data shown in the main paper were taken for 2 nm Ni(111)/W(110) films. We also measured 9 nm Ni(111)/W(110) films (see supplementary material and Fig. S8) to rule out the contributions from superdiffusive spin transfer to the results shown here.

### Time-resolved angle-resolved photoemission spectroscopy (tr-ARPES)

The tr-ARPES measurements were performed at the PG2 beamline of the FLASH free electron laser at DESY/Hamburg using the HEXTOF endstation *(42)*. A momentum microscope with photoelectron time-of-flight analysis measures 3D photoelectron intensity as a function of binding energy, $E$-$E_F$, relative to the Fermi level, for electron momenta ($k_x$, $k_y$) roughly parallel to the sample plane. The photoelectron momentum perpendicular to the sample surface was varied via the selected FLASH photon energy to probe ARPES from the L-W-U (74 eV) and Γ-K (134 eV) Brillouin zone planes shown in Fig. 1C. The instrument provides a momentum resolution of ~0.05 Å$^{-1}$ and covers a $k_{x,y}$-range of 4 Å$^{-1}$ diameter. The momentum component perpendicular to the surface, $k_z$, was calculated considering an inner potential of 13.3 eV for free-electron-like final states *(29)*. The combined photon and photoelectron energy resolution was determined to 200±4 meV as the full-width at half-maximum (FWHM) of a Gaussian function convoluting a Fermi-Dirac distribution fitted to the un-pumped ARPES data for Ni *sp*-bands crossing the Fermi level at the *k*-points indicated in Fig. 2. We used optical laser pulses of 1.2 eV energy and ~100 fs duration to excite the Ni electronic system. The FLASH pulse length was ~160 fs. Pump and probe pulses were incident on the sample collinearly at an angle of 22° relative to the sample surface. The pump and probe focal spot sizes on the sample were determined to 100 × 200 µm$^2$ and 50 ×100 µm$^2$, respectively. P-polarization was used for both pump and probe beams. This enabled us to measure the pump-probe sum frequency assisted photoemission signal *(43)* for monitoring the temporal pump-probe overlap and determine the combined temporal resolution to 227±2 fs (see supplementary information and Figs. S9, S10). The pump fluences used for the shown experiments were determined to be 2.2 and 3.7 mJ/cm$^2$. All measurements shown were performed at room temperature. The ARPES spectra shown here are averaged over equivalent points in the ($k_x$, $k_y$) region of the BZ to improve the signal-to-noise ratio. It is important to point out that that for the selected *k*-points in Fig. 3 and Figs. S2B, S3B it is possible to obtain the population of majority and minority spin 3*d*-states without the need of cumbersome photoelectron spin analysis. This is enabled by their particular energetic position below and above the Fermi level, respectively. At all other *k*-point photoelectron spin analysis would be required which is beyond the scope of this paper.

Theory
Our recently developed non-equilibrium theory of the wavevector-dependent ultrafast electron and lattice dynamics *(23)* has been extended here to explicitly provide the energy transfer between the laser-excited electrons and the lattice. The main features are the phonon branch and wavevector dependence of electron-phonon coupling and an explicit inclusion of anharmonic effects describing phonon-phonon scattering events. We model the non-equilibrium variation of the wavevector-dependent phonon populations *(24)*. Our kinetic theory captures the full transient dynamics of the non-equilibrium phononic populations. The rate of exchange that defines the time evolution of the non-equilibrium energy flow between the electronic system and the different phonon modes after laser excitation is calculated by numerically solving the following rate equations

$$\frac{\partial E_e}{\partial t} = \sum_{q,v} \hbar\omega_v(q)\gamma_v(q,E_e,t)\left[n_v\left(q,E_l^q\right) - n_v(q,E_e)\right] + P(t), \quad (1)$$

$$\frac{\partial E_v^q}{\partial t} = -\hbar\omega_v(q)\gamma_v(q,E_e,t)\left[n_v\left(q,E_l^q\right) - n_v(q,E_e)\right] + \frac{\partial E^{p-p}}{\partial t} \text{ for } q = q_1,\cdots,q_N \quad (2)$$

where $n_v(q,E_l^q)$ is the out-of-equilibrium phonon population of phonon mode $q$ with branch $v$, with $E_l^q$ being the time-dependent amount of energy stored in this particular mode *(24)*. $P(t)$ is the pump laser field that generates the non-equilibrium electronic distribution. $\gamma_v(q,E_e,t)$ is the phonon linewidth due to e-ph scattering, which depends explicitly on the phonon mode, electronic spin degrees of freedom and on the electronic energy, $E_e$. It is therefore a time-dependent quantity. Note that, while the first terms on the right-hand side of Eq. (2) define the energy flow due to e-p interaction, the term, $\frac{\partial E^{p-p}}{\partial t}$, defines the energy flow due to p-p scattering processes and explicitly accounts for the system anharmonicities *(23)*. To obtain a full solution of the non-equilibrium model defined by Eq. (2), we compute all required material-specific quantities using spin-polarized density functional theory and solve Eqs. (1) and (2) numerically *(23)*. The resulting transient phonon populations 1 ps after laser excitation relative to the thermal equilibrium are displayed in Fig. S13 for the L-W-U and Γ-K planes (see Fig. 1B).

**Acknowledgments**

The authors acknowledge FLASH in Hamburg, Germany, for provision of X-ray free-electron laser beamtime and thank the instrument group and facility staff for their assistance as well as Holger Meyer and Sven Gieschen from the University of Hamburg for support of the HEXTOF instrument.

V.S., X.W. and H.A.D. acknowledge support from the Swedish Research Council (VR).

A.Y. acknowledges support from the Carl Trygger Foundation.

P.M., M.B. and P.M.O. acknowledge support by the Swedish Research Council (VR), the K. and A. Wallenberg Foundation (Grant No. 2022.0079), and the Deutsche Forschungsgemeinschaft (DFG, German Research Foundation) via CRC/TRR 227, project ID 328545488 (Project MF). Part of the calculations were enabled by resources provided by the Swedish National Infrastructure for Computing (SNIC) at NSC Linköping, partially funded by VR through Grant Agreement No. 2018-05973.

Funding by the Deutsche Forschungsgemeinschaft (DFG, German Research Foundation) via TRR 173, Project ID 268565370 (Project A02), and CRC 925, Project ID 170620586 (Project B2), as well as by the German Federal Ministry of Education and Research (BMBF) via Projects 05K22UM2 (Mainz) and 05K22FK2 (Kiel) is gratefully acknowledged.

**Author contributions:**
FLASH measurements: MH, DK, XW, AY, NW, LW, FP, SM
*Ab-inito* calculations: PM, MB, PMO
Data analysis: VS
Writing – original draft: VS, HD
Writing – review & editing: all authors

**Competing interests:** Authors declare that they have no competing interests.
**Data and materials availability:** All data are available in the main text or the supplementary materials. Derived data supporting the findings of this study are available at Shokeen, Vishal & Durr, Hermann. (2023) tr-arpes nickel. Zendo. https://doi.org/10.5281/zendo.8429699.

**Figures and Tables**

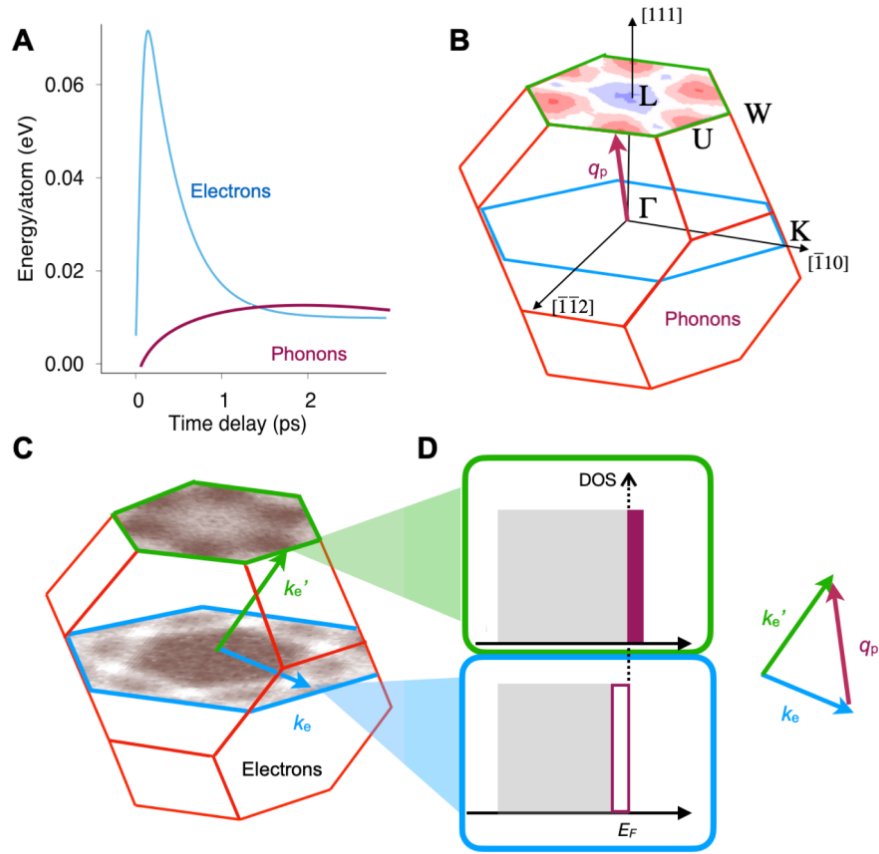

**Fig. 1. Illustration of phonon-electron energy transfer in laser-excited nickel**. (**A**) Energy in the laser-excited electronic system (blue line) is transferred to phonon modes at the Brillouin zone (BZ) boundary (purple line) via electron-phonon (e-p) coupling *(24)*. When the electronic energy becomes less than the energy in a particular phonon mode p-e energy back-transfer can occur. (**B**) Calculated non-equilibrium energy distribution of BZ-boundary phonons (red: more and blue: less; see methods and Fig. S13) compared to equilibrium conditions *(24)*. The purple arrow indicates a phonon momentum vector, $\mathbf{q}_p$, in the BZ of nickel. (**C**) Measured photoemission intensity maps near the Fermi level, $E_F$, in the BZ of Ni within the L-W-U (green border) and Γ-K planes (blue border), respectively. White indicates higher and brown lower intensity. (**D**) Illustration of the change in electron occupation of states upon absorption of phonons with momentum, $\mathbf{q}_p$, and scattering of electrons from states with momentum vector, $\mathbf{k}_e$, to unoccupied states at $\mathbf{k}_e$'. The vector diagram illustrates momentum conservation.

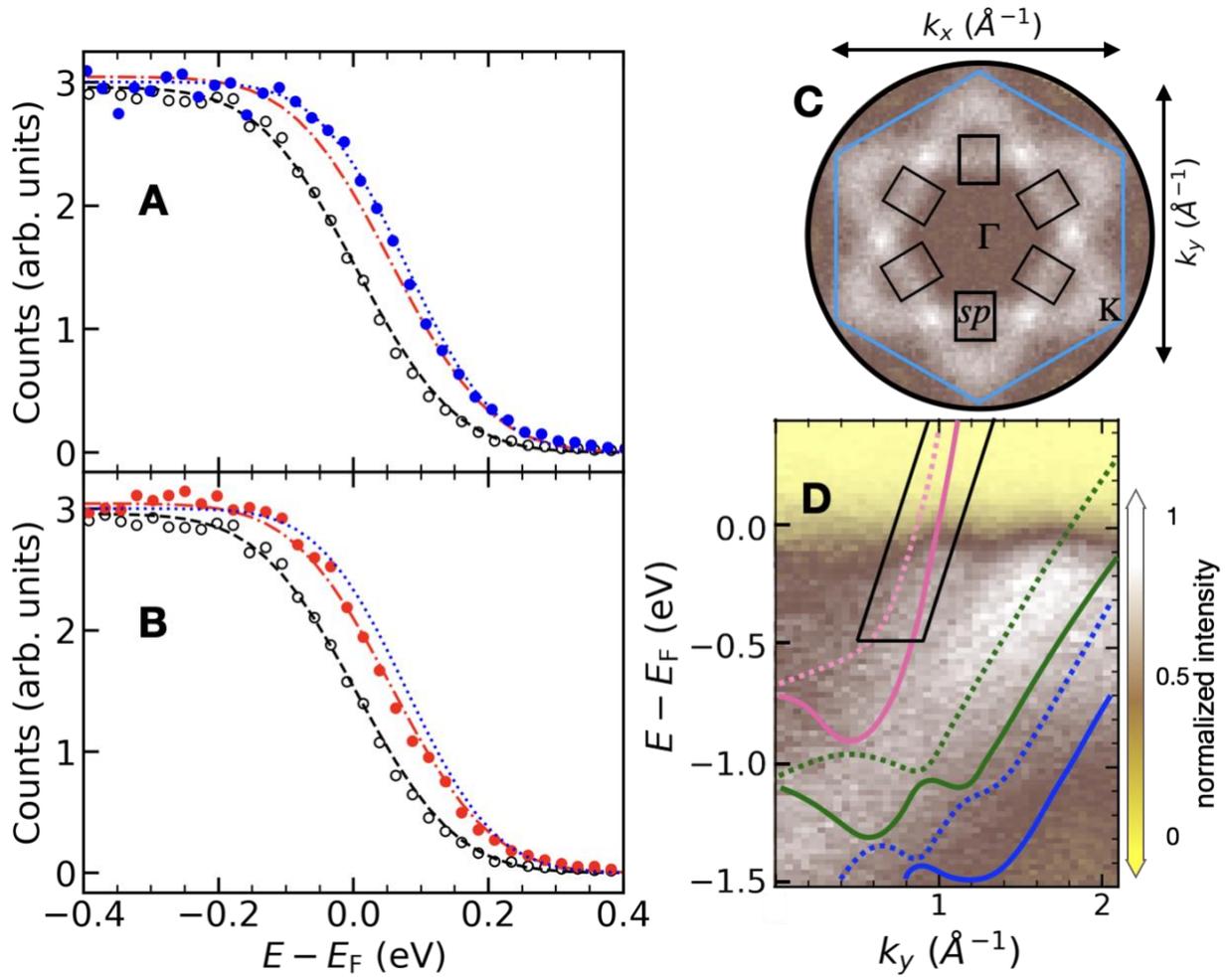

**Fig. 2. Population dynamics of *sp*-bands.** Tr-ARPES data for 2 nm Ni(111)/W(110) obtained for pump fluences of (**A**) 2.2 mJ/cm$^2$ (blue solid circles and lines) and (**B**) 3.7 mJ/cm$^2$ (red solid circles and lines) measured at delay times of 1.8 ps. Open black circles and lines correspond to spectra obtained before arrival of the pump pulses (-0.6 ps time delay). Lines are fits to the data using a Fermi-Dirac distribution function including a chemical potential shift and a constant *sp* density of states across the shown $E$-$E_F$ energy range. (**C**) ARPES data of the integrated intensity near the Fermi level, $E_F$, in the ($k_x$, $k_y$)-plane (see Fig. 1). The black rectangles mark the integration region within which the spectra shown in (**A**, **B**) were obtained. Panel (**D**) shows the ARPES intensity as $E$-$E_F$ vs. $k_y$ along the Γ-K direction (see Fig. 1). The black parallelogram shows the $E$-$E_F$ dependence of the ($k_x$, $k_y$) integration areas (black rectangles) in (**C**).

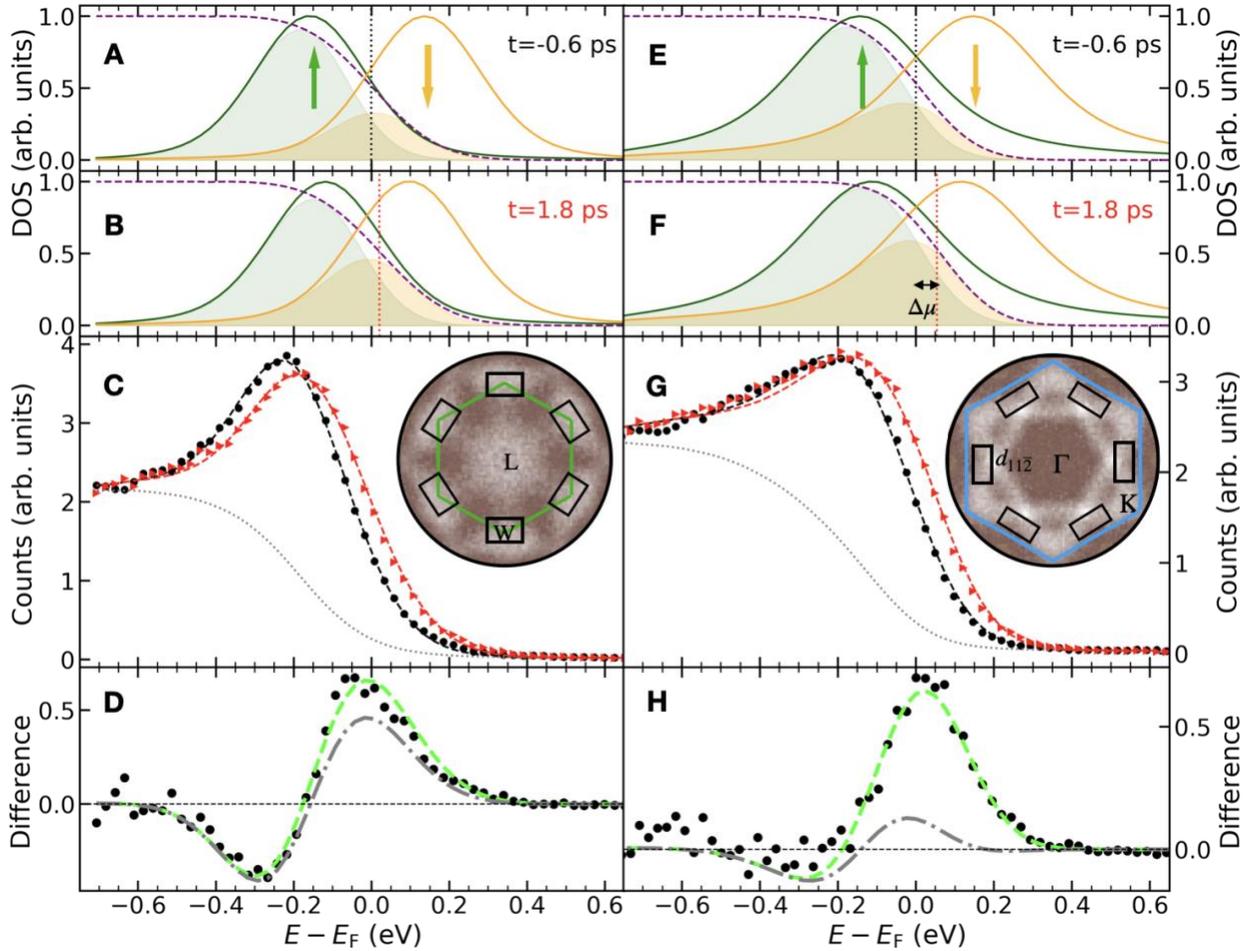

**Fig. 3. Dynamics of 3d-bands at selected k-points.** Tr-ARPES data for 2 nm Ni(111)/W(110) and 3.7 mJ/cm² pump fluence (**A-D**) at the W-point and (**E-H**) at the $d_{11\bar{2}}$ -point. Spectra were obtained at time delays of 1.8 ps and -0.6 ps, *i.e.*, before time-zero, are shown as red and black solid symbols, respectively. Fits of the spectral lineshapes after background (dotted lines in (**C** and **G**)) subtraction *(30)* are shown in (**A** and **E**) and (**B** and **F**) for -0.6 ps and 1.8 ps, respectively. The Lorentzian functions represent the exchange split 3d-bands, the purple dashed lines are the Fermi-Dirac functions and the vertical dotted lines mark the position of the chemical potentials. The resulting fits to the measured spectra are included in (**C** and **G**) as dashed lines. The shaded areas indicate the occupied parts of the respective spin states. (**D** and **H**) show the difference spectra (solid symbols) of the data in (**C** and **G**). The (green) dashed lines are the fits to the data, while the gray dash-dotted lines represent the calculated spectra for zero chemical potential change. The obtained fit parameters are summarized in Fig. 4. The insets in (**C**, **G**) display the measured tr-ARPES intensities around the Fermi level in the respective ($k_x$, $k_y$)-planes of the Brillouin zone and the regions over which the spectra shown in (**C**, **G**) were averaged.

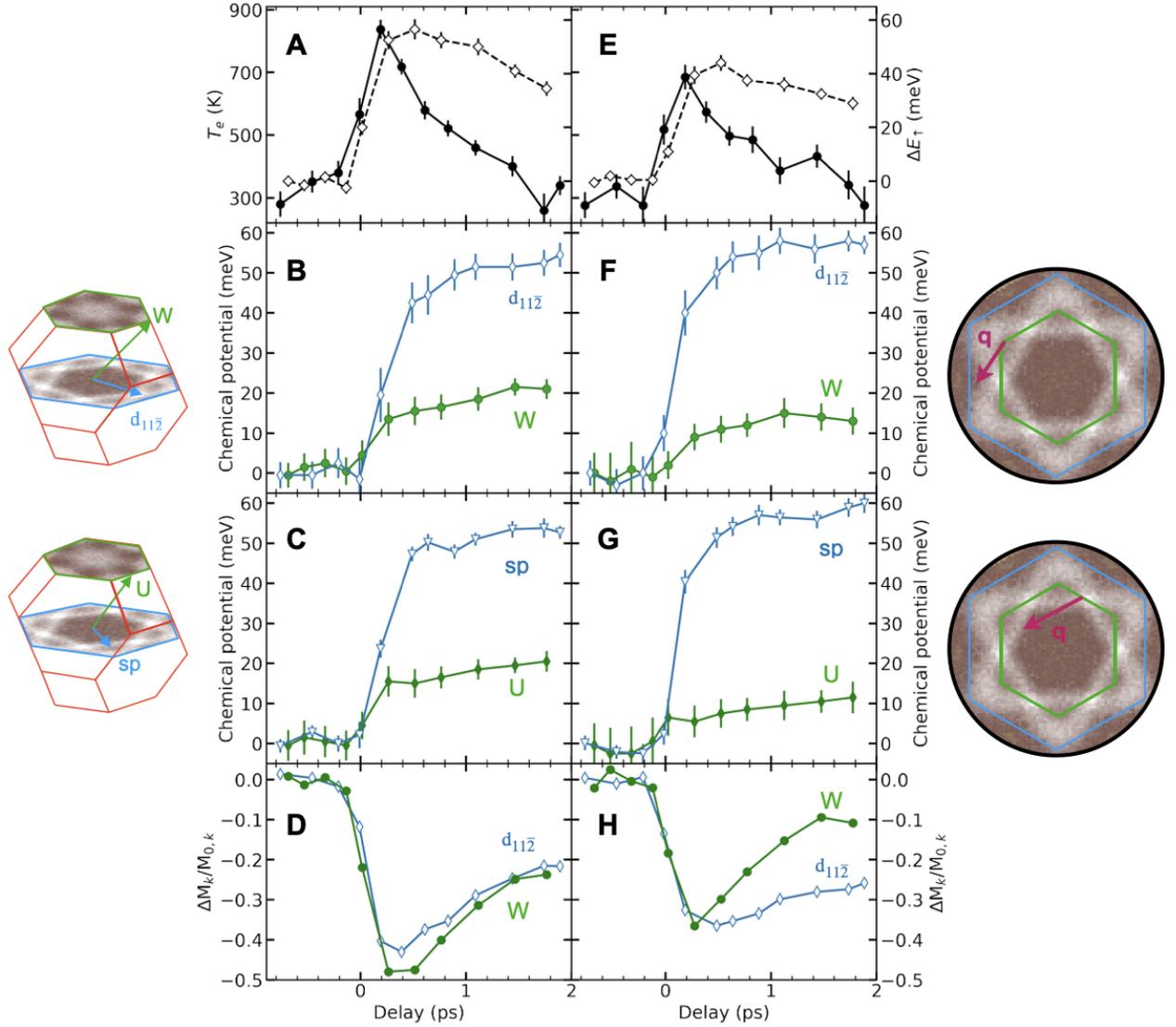

**Fig. 4. Temporal evolution of temperature, magnetization, and chemical potential at selected *k*-points.** (**A**) Laser-induced transient electronic temperature (solid symbols and lines) and average energy shift of the majority 3*d*-bands (see Fig. S4) due to ultrafast demagnetization (open symbols and dashed lines) for a pump fluence of 3.7 mJ/cm$^2$. (**B** and **C**) Chemical potential changes vs. pump-probe time delay at the momentum vectors indicated in the left insets. (D) Magnetic moment change, $M_k$, at the respective k-points. $M_{0,k}$ are the moments before laser excitation. The projection of the electron momentum vectors onto the Γ-K and L-W-U planes are shown in the right insets. The indicated purple wavevectors, **q**, represent BZ-boundary phonon modes. (**E-H**) same as (**A-D**) but for a pump fluence of 2.2 mJ/cm$^2$.

# Supplementary Materials for

**Direct observation of non-equilibrium phonon-electron energy and angular momentum flow in laser-heated Nickel**

Vishal Shokeen, Michael Heber, Dmytro Kutnyakhov, Xiaocui Wang, Alexander Yaroslavtsev, Pablo Maldonado, Marco Berritta, Nils Wind, Lukas Wenthaus, Federico Pressacco, Chul-Hee Min, Matz Nissen, Sanjoy K. Mahatha, Siarhei Dziarzhytski, Peter M. Oppeneer, Kai Rossnagel, Hans-Joachim Elmers, Gerd Schönhense, Hermann A. Dürr

**This PDF file includes:**

    Materials and Methods
    Supplementary Text
    Figs. S1 to S11

## Supplementary text

Time-dependent shift of the valence-band maximum (VBM) in tr-ARPES

Figure S1 shows the momentum dependence of the maximum energy of transient occupied valence states, $E_{VBM}$, defined as the energy where the photoemission intensity near $E_F$ is reduced by half, for selected delay times after optical excitation. In the ground-state (*i.e.,* when the probe pulses arrive at the sample at a delay time, $t$, before the pump, *i.e.* $t < 0$), we find that the highest values of $E_{VBM}$ occur for momenta where real bands are near $E_F$. This occurs when real bands cross the Fermi level, *i.e.*, for *sp*-bands in the Γ-K direction near $k = 1.0$ Å$^{-1}$ (Figs. S1D, H) or for 3*d*-bands in the $\Gamma - d_{11\bar{2}}$ direction near $k = 1.1$ Å$^{-1}$ and 1.6 Å$^{-1}$ in Figs. S1B, F. High values of $E_{VBM}$ also occur near the W-point (Figs. S1A, E) when the majority 3*d*-band approaches $E_F$ without crossing it.

300 fs after laser excitation, $E_{VBM}$ has increased at the W-point by almost 80 meV. For larger delay times $E_{VBM}$ slowly decreases by 30 meV at 1.8 ps. A monotonic increase of $E_{VBM}$ is observed at the *sp*-band Fermi crossing and for 3*d*-states near the $K_{01\bar{1}}$-point. Such a significant momentum dependence of electronic excitations has not been reported before for Ni *(11, 21, 28)* or other ferromagnets *(47, 48)*. The significantly different shift of $E_{VBM}$ observed for different *k*-points excludes space charge effects as a cause of this behavior. For other materials, an increase of $E_{VBM}$ has been attributed to a change of the global chemical potential, *i.e.*, to increased band occupations *(39, 40)*.

Core-level measurements

To corroborate the observed chemical potential shifts in the valence band, we also measured time-resolved 3*p* core level photoemission spectra for the same fluence conditions. The unpumped spectrum in Fig. S7A is averaged between time delays -0.4 ps < t < 0.2 ps. It clearly shows spin-orbit-split peaks whose binding energy was determined by fitting two Voigt functions for each core-level. The background for the core-level spectra has been calculated using the statistics-sensitive non-linear iterative peak-clipping (SNIP) algorithm *(44, 45)*. We obtain values for binding energies of the spin-orbit-split $3p_{1/2}$ and $3p_{3/2}$ peaks of 68.0 and 66.2 eV with a width of 1.7 and 1.8 eV, respectively which agree well with the reports in the literature *(46)*.

Upon pumping at a pump fluence of 3.7 mJ/cm$^2$ we observe a slight attenuation of the amplitude as shown in Fig. S12. This amplitude reduction could be due to enhanced scattering of photoemitted electrons with the increase of phonon population. The change in peak binding energy with pump-probe time delay is illustrated in Fig. S7B for a time-delay of 1.8 ps. In order to reduce the number of fit parameters we keep the ratio of peak amplitudes; peak width and spin-orbit splitting the same at all time delays.

Figure S7B also shows the comparison of the measured difference spectra with two fit conditions which are 2% (1.7%) reduction in amplitude and 12 meV (15 meV) binding-energy shifts towards the Fermi level. The change in the amplitude and peak shift have different signatures in the difference photoemission spectra with absorptive and dispersive behavior respectively. As shown in Fig. S7B, 15 meV shift and 1.7% amplitude decrease results in better fit (gray plot) in the slope around 66 eV but introduces a positive feature around 65 eV whereas blue fits better at 65 eV at the cost of poor fit in the slope.

Correction of the Light Assisted Photoemission Effect (LAPE)
The LAPE is a transient process occurring only during the temporal overlap of intense infra-red (IR) pump and XUV probe *(43)*. LAPE appears as a step in the photoemission intensity near Fermi level and is described as the laser (IR) dressing of the emitted photoelectrons in the continuum modifying the photoemission spectrum as shown in Fig S9. Above the Fermi level additional photoemission intensity appears due to the absorption of an IR photon by an XUV emitted photoelectron. The above-Fermi intensity distribution roughly follows the direct photoemission spectrum at lower binding energy shifted by the laser photon energy of 1.2 eV. The change in intensity above Fermi level (red line in Fig. S9A) relative to the ground-state photoemission spectra (dashed black lines) is a sensitive measure of the temporal duration of pump and probe pulses (see Fig. S10). We obtain values of 227±2 fs for the convoluted pump and probe pulse duration. For analyzing the spectral lineshape of photoemission spectra in temporal pump-probe overlap it is necessary to subtract the LAPE induced lineshape changes. Since we only analyze the lineshape in the vicinity of the Fermi level it is possible to remove the LAPE modifications as indicated in Fig. S9B. The LAPE corrected spectrum shown in red in Fig. 9B is very similar to the unpumped spectrum justifying the correction procedure in the binding energy range of ±1 eV around the Fermi level where significant effects due to pump laser are expected to occur. This correction procedure is then repeated for other time delays where LAPE is present (i.e. for delay times between -0.2 ps < t < 0.2 ps) to extract the pump induced dynamical signal.

Fitting valence photoemission spectra
We used the fit functions described in Eqs. (S1-S3) for modelling the photoemission signal for k-dependent density of states. The fit function generally includes the product of density of states with Fermi-Dirac distribution convoluted with a gaussian function in Eq. (S1). The Fermi-Dirac distribution determines the filling of the density of states while the gaussian convolution mimics the instrumental resolution. The evolution of the Fermi function for different k-points provides information about the change in the chemical potential and electronic temperature induced broadening around the Fermi level. The measured photoemission intensity is then

$$I(E') = \left(f_{dos}(E') * f_{FD}(E',\mu,T_e)\right) \otimes G(E',\delta) \tag{S1}$$

where $E' = E - E_F$, is the electron energy relative to the Fermi level, $E_F$. $f_{dos}$, $f_D$, $\mu$ and $T_e$ represent the density of states, Fermi-Dirac distribution, chemical potential and electronic temperature respectively. $\otimes$ represents the convolution of a gaussian function, $G$, with $\delta$ as its FWHM describing the energy resolution of the experiment.

To extract the dynamics of exchange splitting from *k*-points with minority and majority *d*-bands, the density of states for up and down spin *d*-bands is produced by using Lorentzian functions Eq. (S2) as

$$f_{d-dos} = l_{min}(E', x_0 + E_{ex}, dx) + l_{maj}(E', x_0, dx) \tag{S2}$$

with $x_0, E_{ex}$ and $dx$ representing the energy position, exchange splitting and width of Lorentzian functions for minority and majority *d*-bands. The electronic temperature for these k-points are the values extracted from the fitting of *sp*-bands. The delay-independent background for *d*-bands is extracted using the Shirley method (*30*). The extracted energy shift of majority band for different *k*-points has been evaluated using two exponential functions convolved with temporal resolution ($\delta_t$) shown in Eq. (S3) as

$$\Delta E_{maj} = a * \left( \exp\left(-\frac{t}{\tau_{demag}}\right) - \exp\left(-\frac{t}{\tau_{remag}}\right) \right) \otimes \text{erf}(\delta_t) \tag{S3}$$

where $\tau_{demag}$ and $\tau_{remag}$ are demagnetization and remagnetization time constants, respectively, erf is the error function.

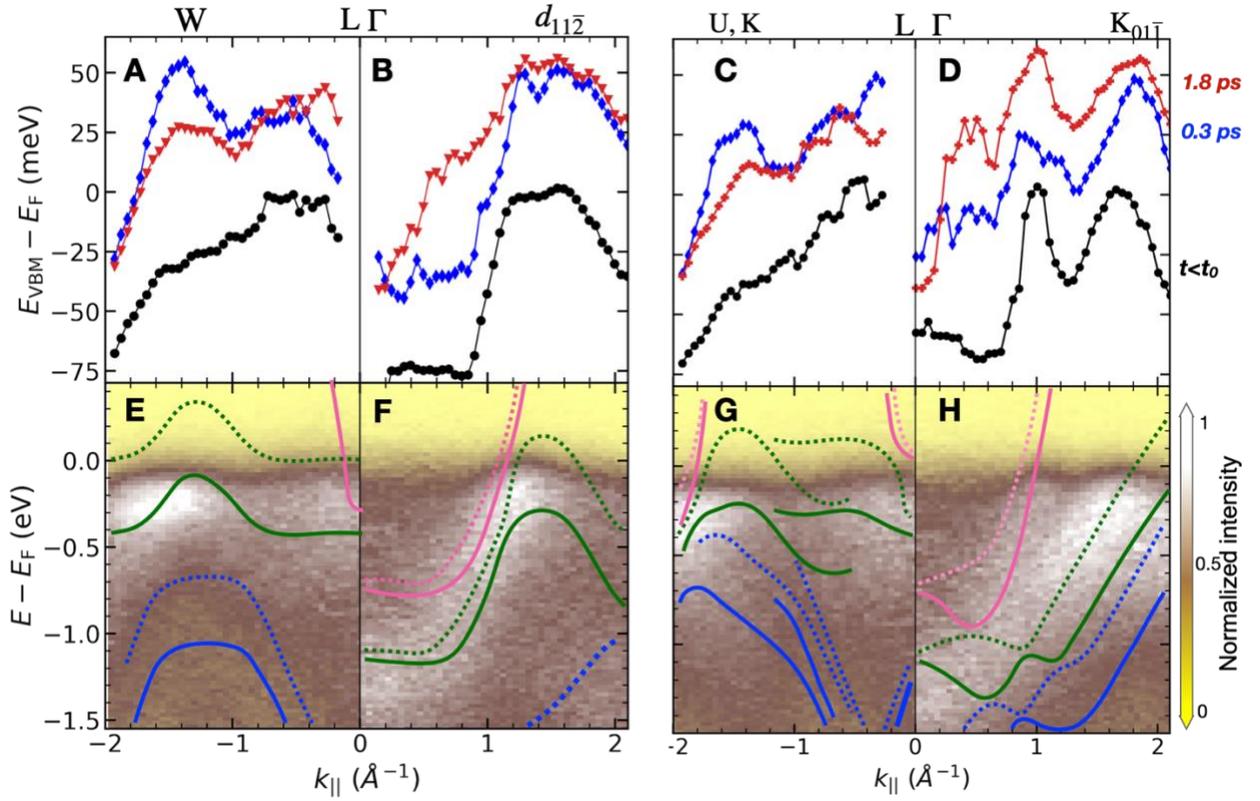

**Fig. S1. Changes in the photoemission signal from states near the Fermi level, $E_F$.** (**A-D**) Energy position of the top-most occupied valence-band electronic state, $E_{VBM}$, relative to the Fermi level, $E_F$, along the respective directions for various pump-probe time delays. $E_{VBM}$ was chosen as the inflection point of data such as the ones shown in Fig. 2. (**E-H**) Momentum resolved ARPES intensity (color shading) and calculated electronic state dispersions (lines) along L-W, $\Gamma$-$d_{11\bar{2}}$, L-U and $\Gamma$-K Brillouin-zone directions (see Fig. 1C), respectively. Data were obtained for a pump fluence of 3.7 mJ/cm$^2$.

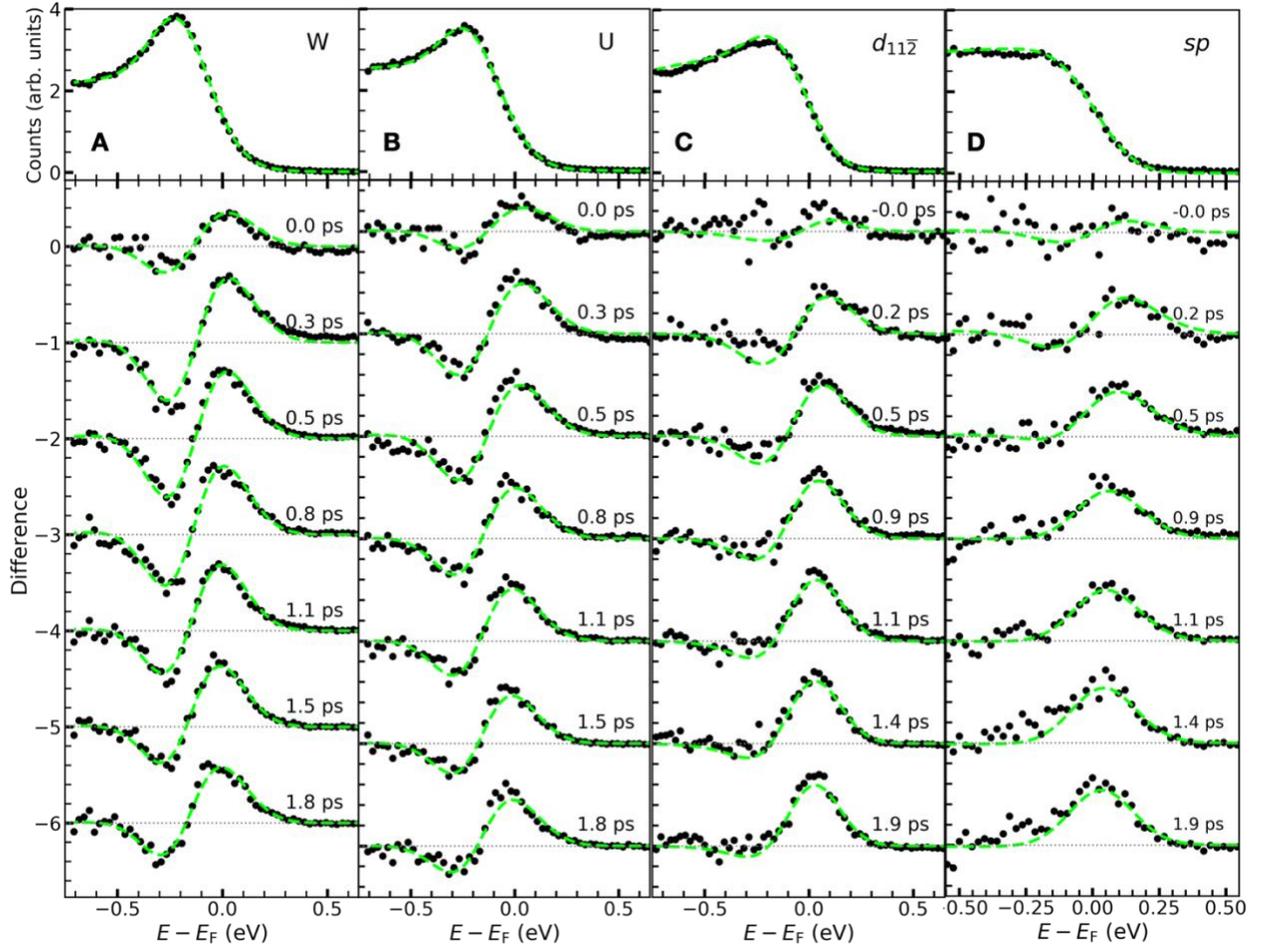

**Fig. S2. Tr-ARPES from *k*-points at 3.7 mJ/cm² pump fluence.** Shown are the ARPES data at (**A**) W, (**B**) U, (**C**) $d_{11\bar{2}}$ and (**D**) *sp* points obtained before laser pump (top panels) as well as difference spectra at the indicated pump-probe time delays integrated over ±150 fs (waterfall plots in bottom panels) for the indicated k-points averaged over the k-regions shown in Fig. S1. Data were obtained for a pump fluence of 3.7 mJ/cm². Regions in ($k_x$, $k_y$) over which the data of this figure were averaged are indicated in Fig. S5.

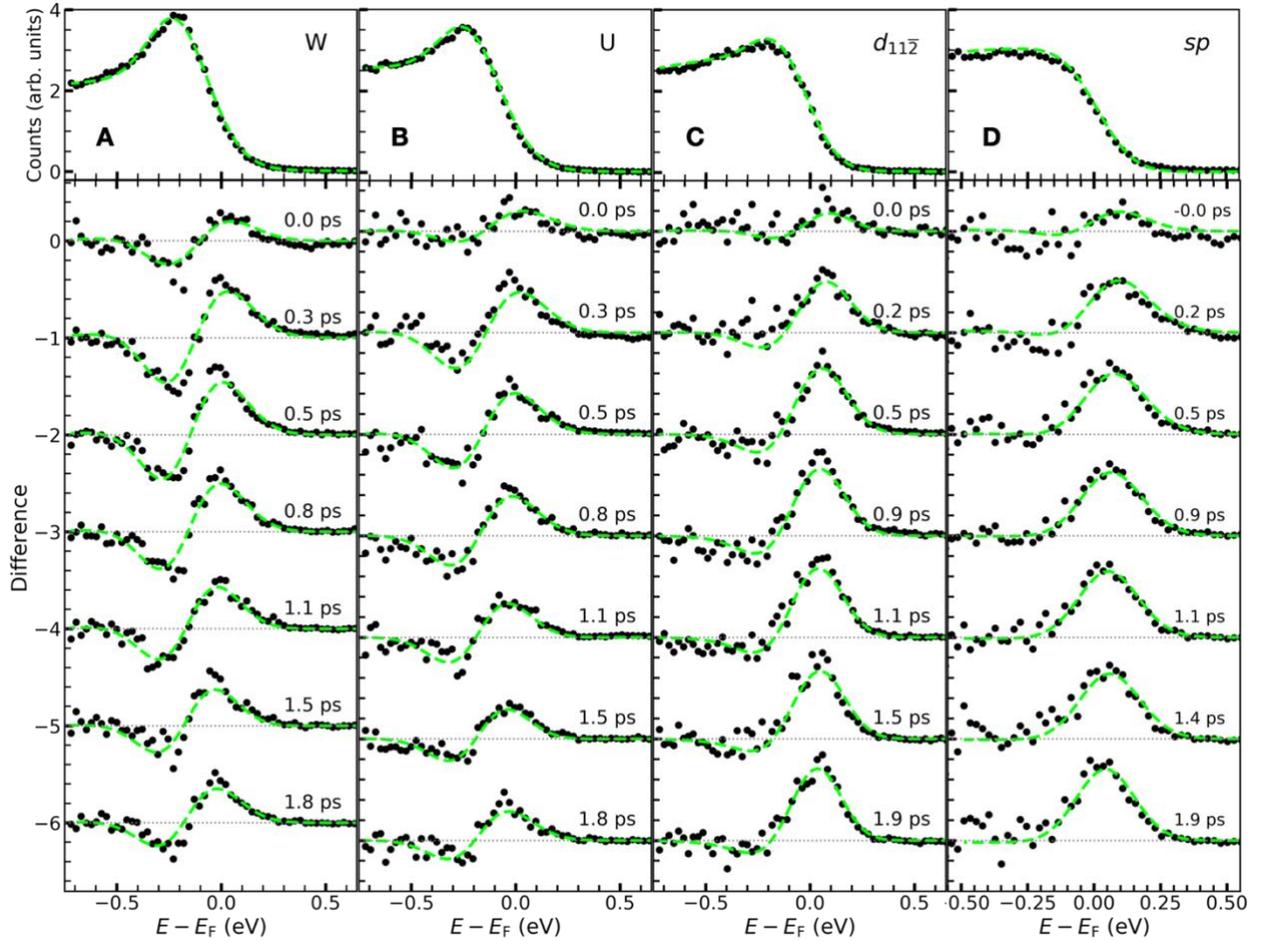

**Fig. S3. Tr-ARPES from *k*-points at 2.2 mJ/cm² pump fluence.** Shown are the ARPES data obtained at (**A**) W, (**B**) U, (**C**) $d_{11\bar{2}}$ and (**D**) *sp* points before laser pump (top panels) as well as difference spectra at the indicated pump-probe time delays integrated over ±150 fs (waterfall plots in bottom panels) for the indicated k-points averaged over the k-regions shown in Fig. S1. Data were obtained for a pump fluence of 2.2 mJ/cm². Regions in ($k_x$, $k_y$) over which the data of this figure were averaged are indicated in Fig. S5.

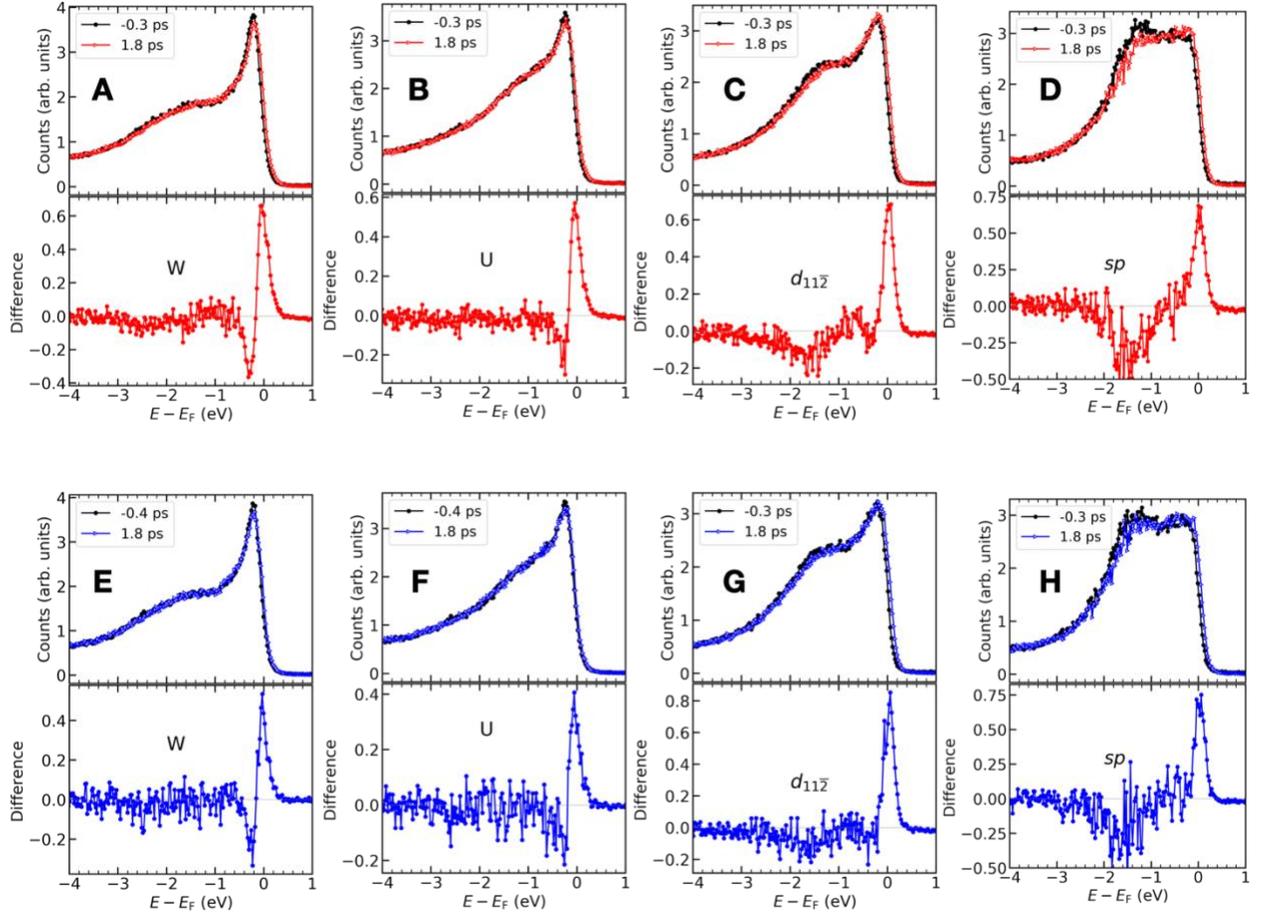

**Fig. S4. Tr-ARPES from W, U, $d_{11\bar{2}}$ and sp-points for a broader BE range from 1 to -4 eV around $E_F$.** (**A-D**) Pump fluence 3.7 mJ/cm$^2$, (**E-H**) 2.2 mJ/cm$^2$. The black lines and symbols represent the energy spectra for t < -0.3 ps and colored (red, blue) lines and symbols are for a delay of 1.8 ps. All data were averaged over ±300 fs. Regions in ($k_x$, $k_y$) over which the data of this figure were averaged are indicated in Fig. S5.

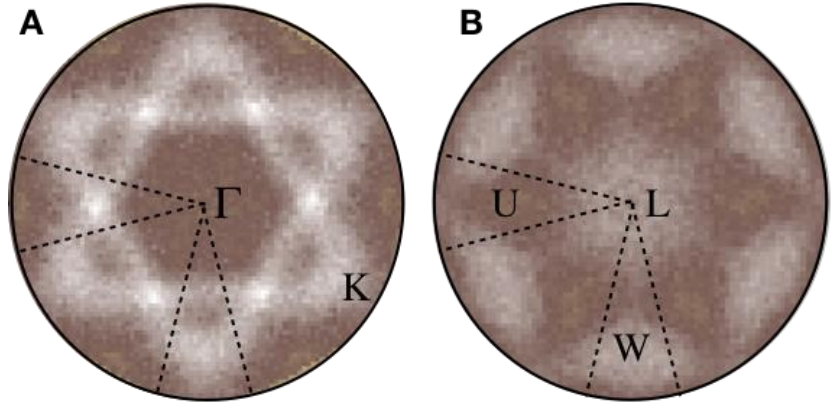

**Fig. S5. ARPES from states near the Fermi level in Ni(111).** (**A**) photoemission intensity from states near E$_F$ in the Γ-K plane of the Ni(111) Brillouin zone (see Fig. 1C) (**B**) same as in (**A**) but for the L-U-W plane in Fig. 1C. The shown regions bounded by dotted lines along different *k*-directions are used for the analysis in Fig. 2 and Fig S2.

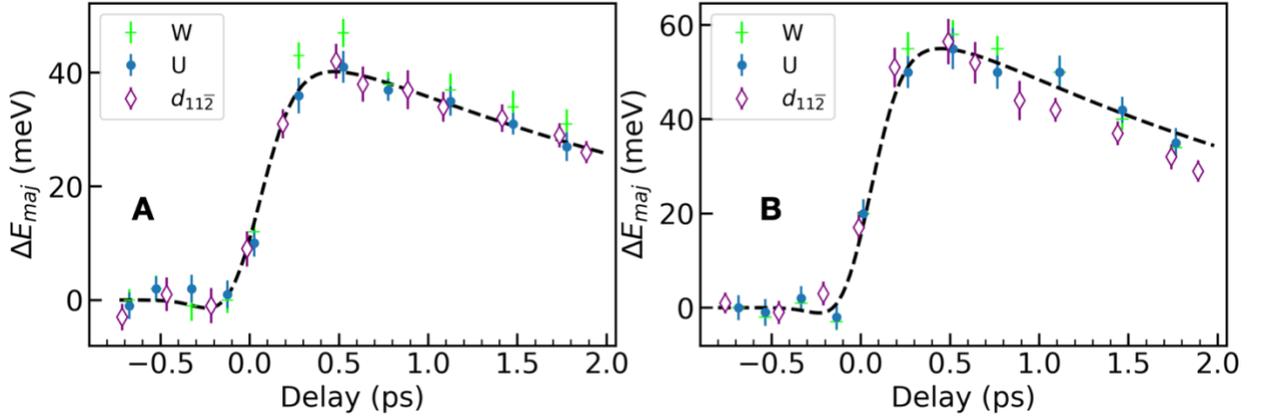

**Fig. S6. Temporal evolution of the 3*d* exchange splitting at different *k*-points.** Energy positions, Δ$E_{maj}$, of the majority spin-bands for tr-ARPES spectra at the W, U and $d_{11\bar{2}}$-points obtained by fits as shown in Fig. 4 of the main paper for pump fluences of (**A**) 2.2 mJ/cm$^2$ and (**B**) 3.7 mJ/cm$^2$. Dashed black lines represent the fits to the data according to Eq. (S3) with demagnetization times, $\tau_{demag}$, of 205±50 fs and 200±70 fs, and remagnetization times, $\tau_{remag}$, of 3.0±0.5 and 2.9±0.7 ps for 2.2 and 3.7 mJ/cm$^2$, respectively.

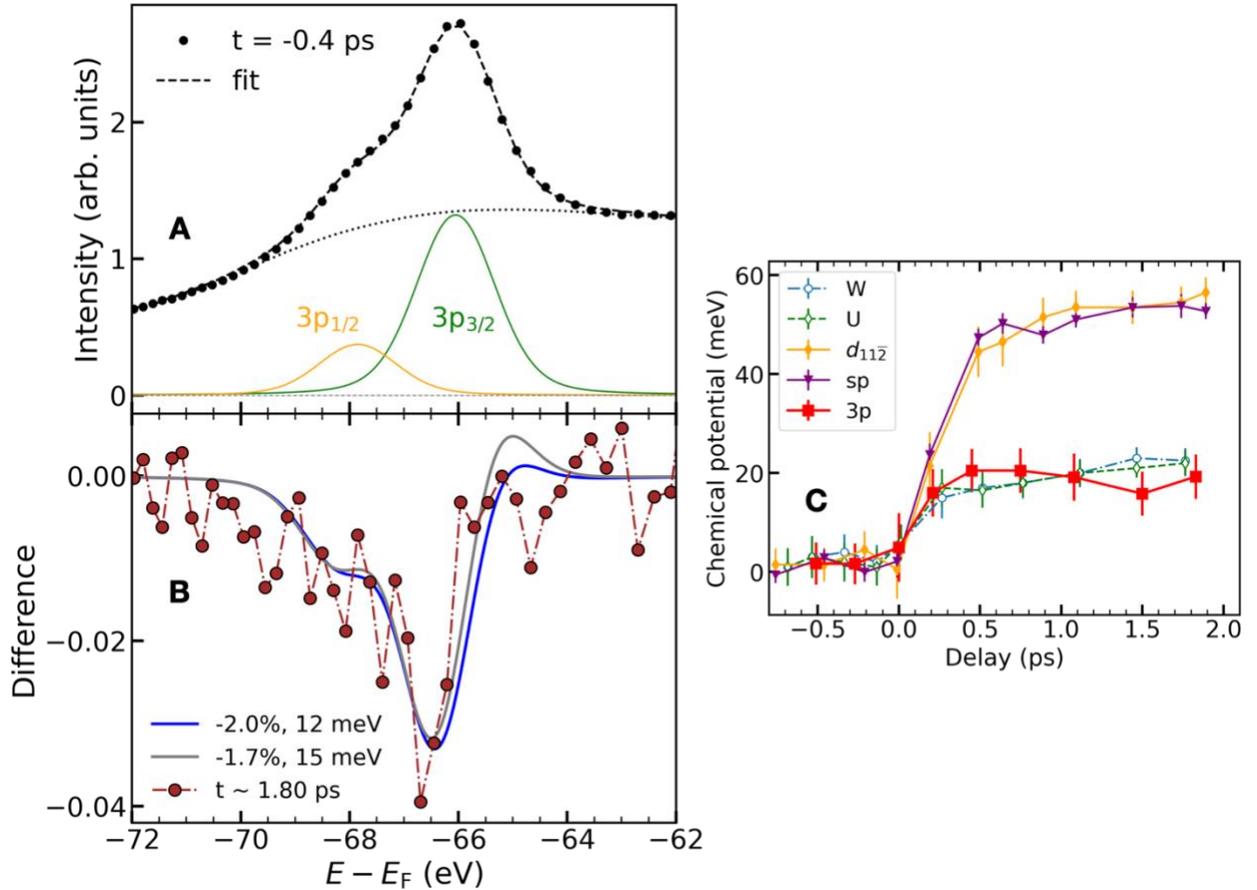

**Fig. S7. Temporal evolution of Ni(111) 3p core-level photoemission**. (**A**) 3p core-level photoemission spectra measured with 204 eV shown in black markers. The dotted black line is the calculated background using the SNIP method described in the supplementary text. The dashed line is the fit obtained using two Voigt functions representing the spin-orbit-split $3p_{1/2}$ and $3p_{3/2}$ core levels (colored solid lines). (**B**) Difference between pumped and unpumped core-level photoemission spectra at a pump fluence of 3.7 mJ/cm² (dash-dotted line and circles). Blue and gray lines are the fits obtained for two conditions as described in the supplementary text. (**C**) Temporal evolution of the core-level-shift (red line and squares) towards the Fermi level compared with chemical potential change for the indicated $k$-points.

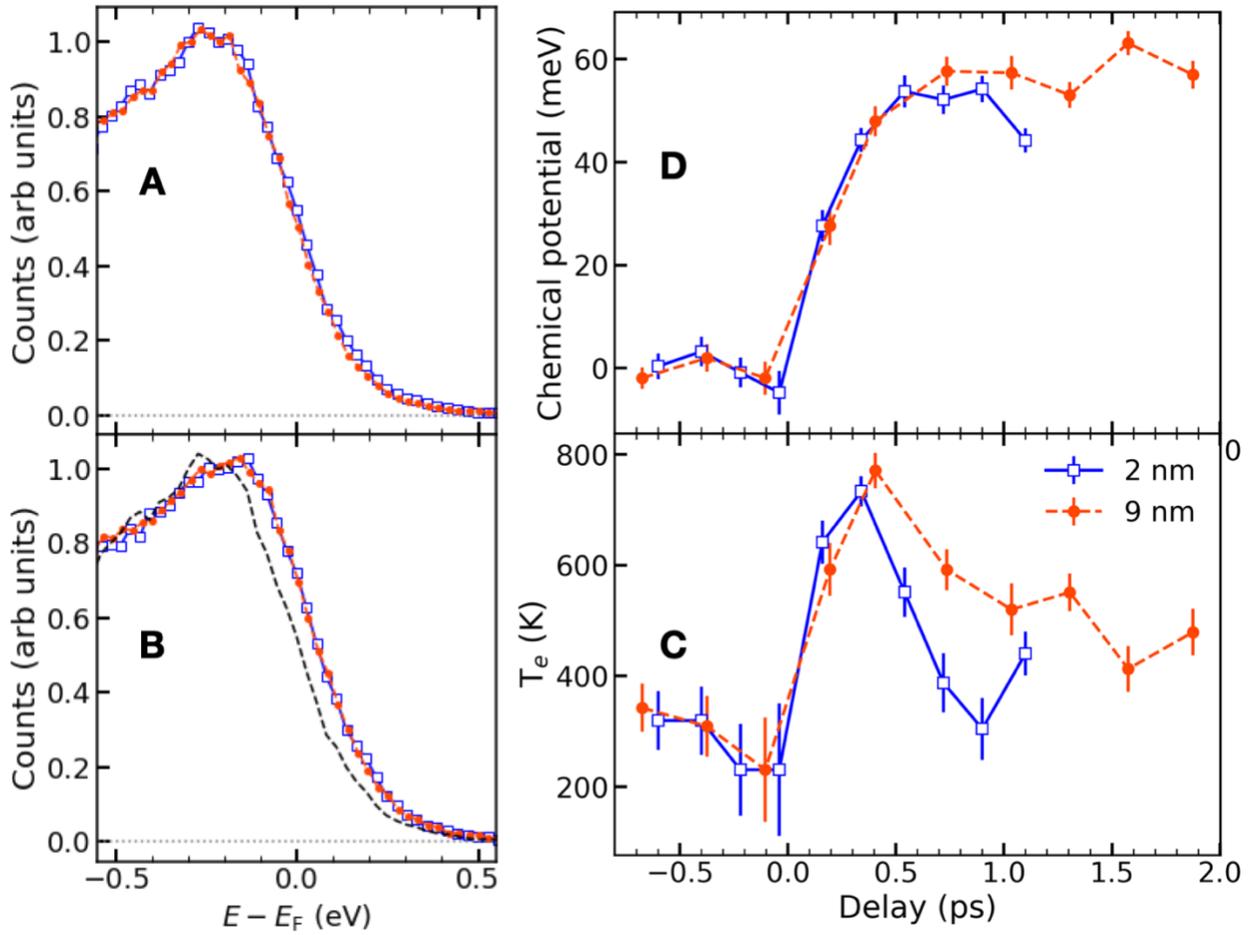

**Fig. S8. Tr-ARPES of Ni(111)/W(110) of different Ni thicknesses.** (**A**) Unpumped angle-integrated photoemission spectra taken for Ni films of 2 nm thickness (blue lines and open squares) and 9 nm thickness at 205 eV photon energy (red lines and solid circles). At 205 eV photon energy similar parts of the Ni BZ are probed as at 74 eV. (**B**) Pumped angle-integrated photoemission spectra taken for Ni films of 2 and 9 nm thickness as in (**A**) averaged over ±0.3 ps around a pump-probe time delay of 1 ps. (**C**) Measured electron temperatures, $T_e$, vs. delay time for the 2 and 9 nm thick Ni films. The pump fluence of the 2 nm film was 3.7 mJ/cm² as in the main paper. The pump fluence of the 9 nm film was adjusted to achieve a similar maximum $T_e$ value, i.e. a similar deposited pump laser energy. (**D**) Chemical potential change for the two Ni thicknesses with pump-probe delay.

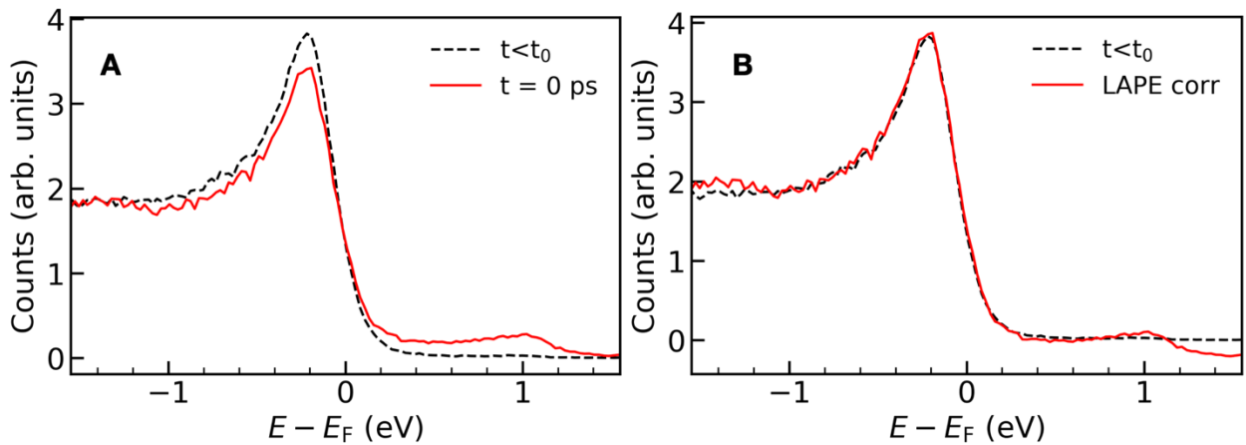

**Fig. S9. Correction of the Light Assisted Photoemission Effect (LAPE). A**) The black dashed line is the unpumped photoemission spectrum and the red solid line is the pumped photoemission spectrum with pump and probe pulses in temporal overlap. The deviation of both spectra is due to the LAPE effect. (**B**) The pumped spectrum after LAPE correction (solid red line) is compared with black unpumped spectrum (dashed black line). LAPE correction is done by first subtracting an offset to remove the step above the Fermi level and then multiplying the remaining spectrum by a factor to match it to the unpumped spectrum.

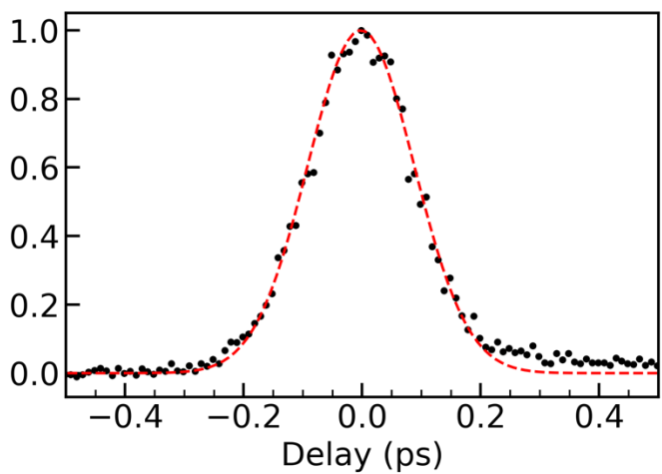

**Fig. S10. Temporal instrumental resolution.** The black markers represent the photoemission signal above the Fermi level in the LAPE region (see Fig. S9 A). The red dashed line is a gaussian fit with a FWHM of 227±2 fs.

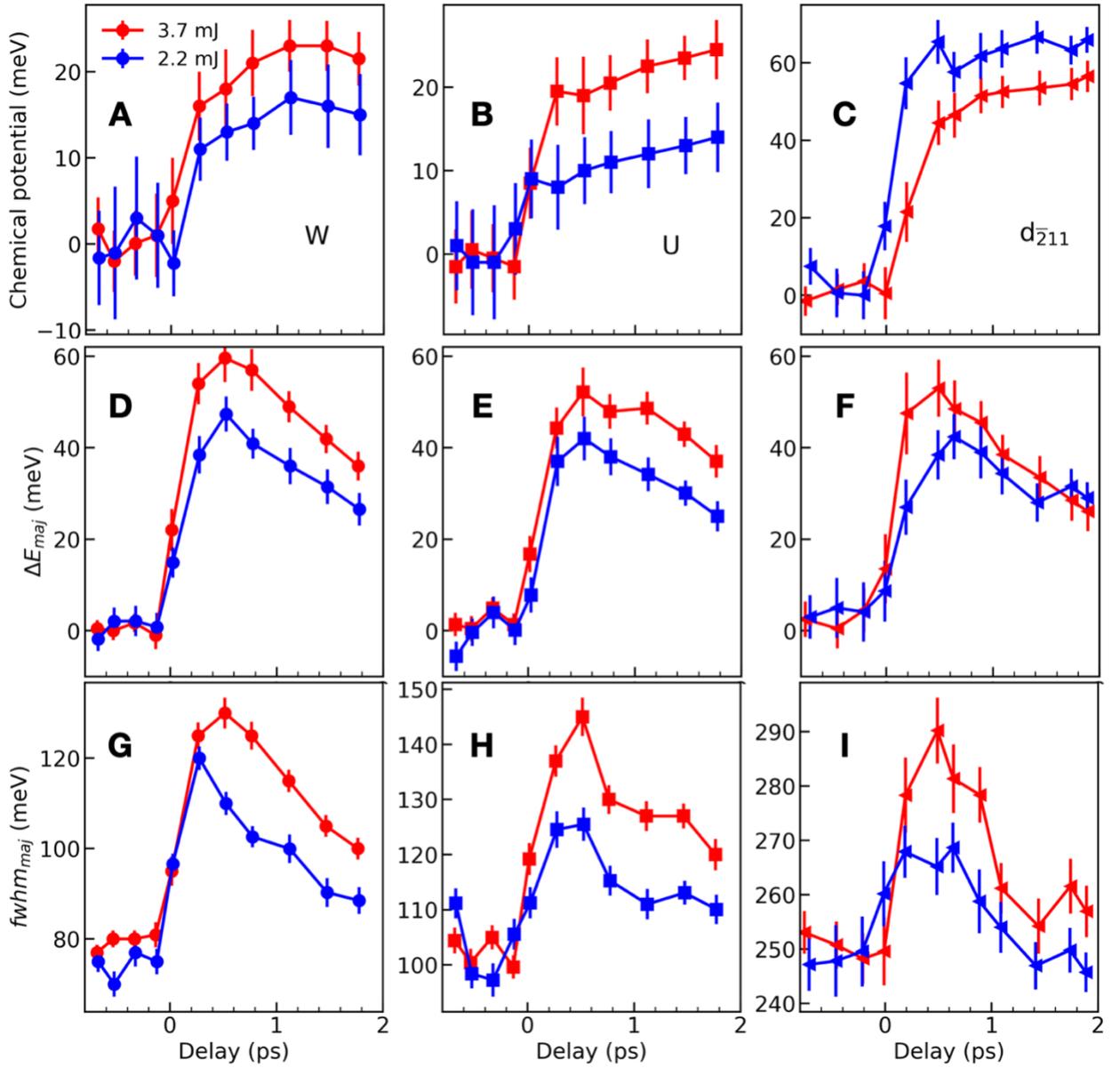

**Fig. S11. Extracted fit parameters at W, U and $d_{11\bar{2}}$ points with exchange split d-bands for 3.7 (red) and 2.2 (blue) mJ/cm$^2$ pump fluence.** (**A-C**) chemical potential shift, (**D-F**) shift of majority band due to exchange collapse. (**G-I**) full width half maximum (FWHM) broadening of majority bands due to laser excitation as a function of pump probe delay.

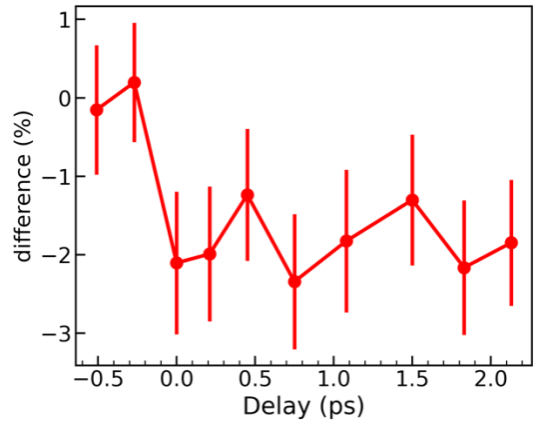

**Fig. S12. Laser induced reduction of 3p peaks.** This figure shows the relative change in the mean amplitude of the 3p peaks for different pump-probe delays. There is a reduction in the amplitude of 3p peaks by ~1.7-2% after laser excitation.

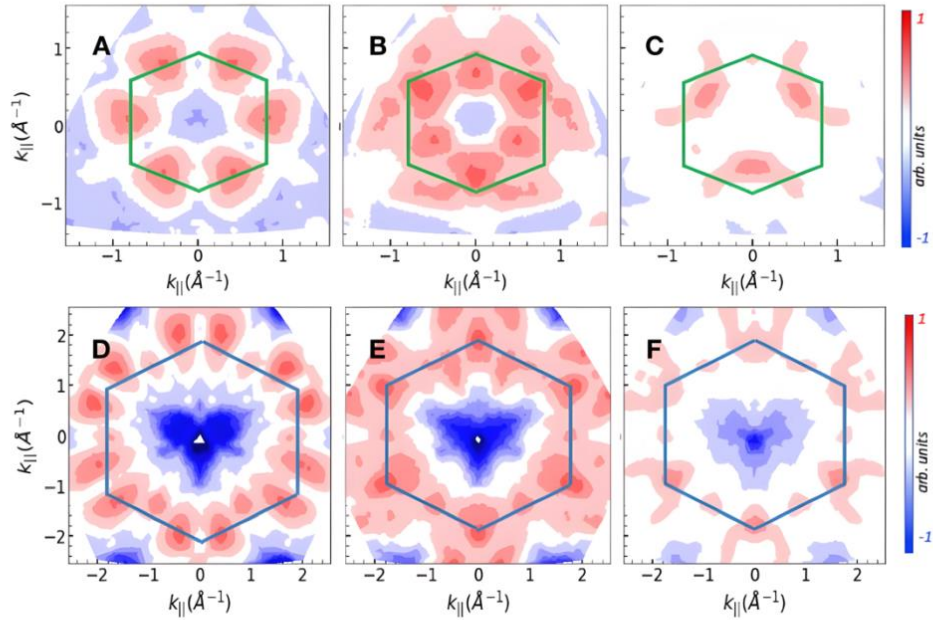

**Fig. S13. Calculated transient phonon occupations.** This figure shows the differences of transient phonon occupation relative to those in thermal equilibrium 1 ps after fs laser excitation in the L-W-U (**A-C**) and Γ-K (**D-F**) planes of the Ni Brillouin zone (see Fig. 1B). Panels (**A, B, D, E**) show phonons with transverse polarization while (**C, F**) display longitudinal phonon polarization.

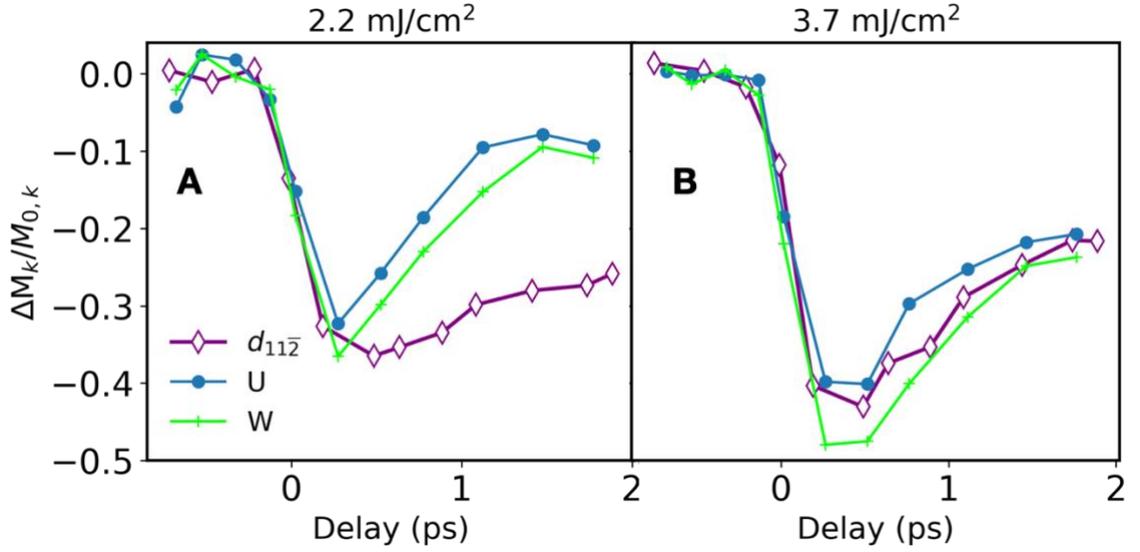

**Fig. S14. Momentum-dependent magnetic moment evolution.** This figure shows the changes of the magnetic moments at the respective $k$-points vs. pump-probe time delay for pump fluences of (**A**) 2.2 mJ/cm² and (**B**) 3.7 mJ/cm². $\frac{\Delta M_k}{M_k}$ is defined as $\frac{\Delta M_k}{M_k} = \frac{N_{maj} - N_{min}}{N_{0,maj} - N_{0,min}}$, where $N_{maj,min}$ are the occupations on majority and minority $d$-bands determined from Fig. 3 and $N_0$ represents the respective occupation numbers before laser excitation.